\documentclass[aip,jcp,reprint,superscriptaddress,letterpaper,floatfix]{revtex4-2}

\usepackage{amsmath}
\usepackage{amssymb}
\usepackage{braket}
\usepackage{mathtools}
\usepackage{bm}
\usepackage {verbatim}

\usepackage{hyperref}
\hypersetup{
    colorlinks=true,
    linkcolor=black, % color for internal links
    citecolor=black, % color for citations
    filecolor=black, % color for file links
    urlcolor=black% color for URL links
}
\usepackage[table]{xcolor} % Enable colors in tables
\usepackage{colortbl}

\begin{document}
\title {Enhanced second-harmonic generation from WS$_2$/ReSe$_2$ heterostructure}

\author{Kanchan Shaikh}
\thanks{These authors contributed equally to this work.}
\affiliation 
{Department of Chemistry, University of Michigan, Ann Arbor, MI 48109, USA}
\author{Taejun Yoo}
\thanks{These authors contributed equally to this work.}
\affiliation 
{Department of Chemistry, University of Michigan, Ann Arbor, MI 48109, USA}
\author{Zeyuan Zhu}
\thanks{These authors contributed equally to this work.}
\affiliation 
{Department of Chemistry, University of Michigan, Ann Arbor, MI 48109, USA}
\author{Qiuyang Li}
\affiliation{Department of Physics, University of Michigan, Ann Arbor, MI 48109, USA}
\author{Amalya C. Johnson}
\affiliation{Department of Materials Science and Engineering, Stanford University, Stanford, CA 94305, USA}
\author{Hui Deng}
\affiliation{Department of Physics, University of Michigan, Ann Arbor, MI 48109, USA}
\affiliation{Department of Electrical Engineering and Computer Science, University of Michigan, Ann Arbor, MI 48109, USA}
\author{Fang Liu}
\affiliation {Department of Chemistry, Stanford University, Stanford, CA 94305, USA}
\author{Yuki Kobayashi}
\email{ykb@umich.edu}
\affiliation {Department of Chemistry, University of Michigan, Ann Arbor, MI 48109, USA}

\date{\today}

%%%%%%%%%%%%%%%%%%%%%%%%%%%%%%%%%%%%%%%%%%%%%%%%%%%%%%%%%%%%%%%%%%%%%

\begin{abstract}
Van der Waals stacking presents new opportunities for nonlinear optics with its remarkable tunability and scalability.
However, the fundamental role of interlayer interactions in modifying the overall nonlinear optical susceptibilities remains elusive.
In this letter, we report an anisotropic enhancement of second-harmonic generation (SHG) from a WS$_2$/ReSe$_2$ heterobilayer, where the individual composite layers possess distinctive crystal phases. 
We investigate polarization-resolved response and twist-angle dependence in SHG and reveal that band alignment alone is insufficient to explain the observed anisotropy in the modified SHG response.
Spectral shifts in excitonic features highlight band renormalization, supporting the role of hybridization between the two layers.
Furthermore, SHG enhancement is highly anisotropic and can even be suppressed in some orientations, suggesting possible intensity-borrowing mechanisms within the heterostructure.
Our work demonstrates the ability to tune both the intensity and polarization dependence of nonlinear optical responses with van der Waals stacking of distinctive crystal phases.
\end{abstract}

%%%%%%%%%%%%%%%%%%%%%%%%%%%%%%%%%%%%%%%%%%%%%%%%%%%%%%%%%%%%%%%%%%%%%

\maketitle

\section{Introduction}

Van der Waals stacking of two-dimensional materials, such as graphene and transition-metal dichalcogenides (TMDs), has led to discoveries of several emergent quantum phenomena.\cite{geim_van_2013, castellanos_van_2022} 
Recent experimental reports include superconductivity in twisted bilayer graphene,\cite{cao_unconventional_2018} formation of Moiré bands in twisted bilayer TMDs,\cite{molino_influence_2023} spin and valley polarization,\cite{hsu_optically_2015, ye_electrical_2016} and the manipulation of interlayer excitons.\cite{reho_excitonic_2024}
With rapid advances in fabrication technologies, two-dimensional van der Waals heterostructures are envisioned as compact and tunable platforms for optics and electronics.\cite{Jia22,Kireev24}

One of the important properties that can be engineered by van der Waals stacking is nonlinear optical response.
In particular, second-harmonic generation (SHG), as a fundamental nonlinear optical process, has been attracting wide attention in two-dimensional material research.
A recent study has shown that a heterostructure of bilayer MoS$_2$ and monolayer graphene, both of which lack broken-inversion symmetry, can exhibit efficient and tunable SHG.\cite{zhang2023emergent}
In another study, a heterostructure was fabricated using monolayer MoS$_2$ and monolayer alloy MoS$_{2x}$Se$_{2(1-x)}$, and enhancement of SHG due to interlayer coupling was revealed.\cite{le_effects_2020}
These previous studies have pointed to important roles of elusive interlayer effects such as interlayer charge transfer, \cite{zhang2023emergent} which breaks inversion symmetry and induces second-order hyper-polarizability; and interlayer coupling, \cite{le_effects_2020} which drives band offset and electronic band hybridization between two constituent layers.
Nevertheless, the elusive role of interlayer effects underlying SHG enhancement in van der Waals heterostructures have yet to be fully understood.

In this work, we show that van der Waals stacking of monolayer $1H$-WS$_2$, which is SHG-active, \cite{janisch_extraordinary_2014} and monolayer $1T'$-ReSe$_2$, which is nearly SHG-inactive, \cite{kucukoz_boosting_2022} can induce up to a $101\%$ increase in SHG response compared to monolayer WS$_2$ alone.
We investigate polarization dependence and twist-angle dependence, and the results show that the stacking drastically affects the anisotropic response while total SHG counts increase consistently for smaller twist angles.
With additional consideration of the band alignment and excitonic features, the observed enhancement is attributed to induced out-of-plane polarization and interlayer band renormalization.

\begin{figure*}[tb]
    \centering
    \includegraphics[scale=1]{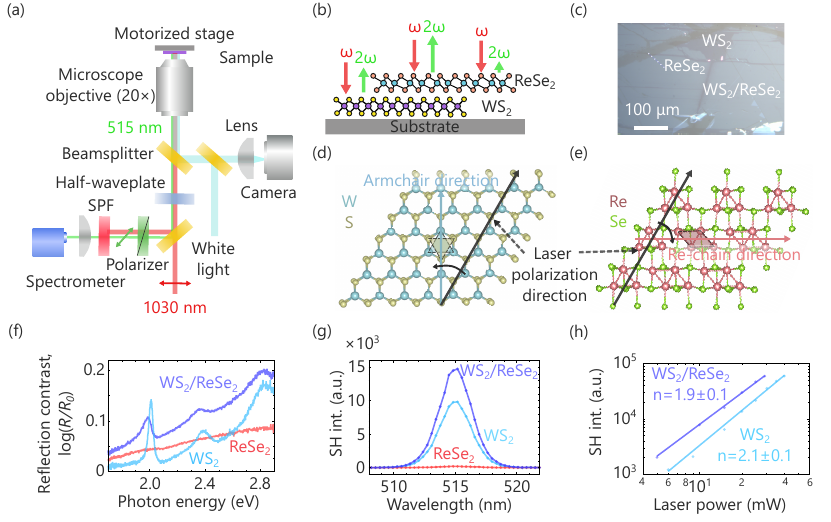}
    \caption{\textbf{Outline of experiment and sample characterization.}
    (a) Sketch of the laser microscopy beamline for the SHG experiments. SPF: Short-pass filter.
    (b) Schematic of SHG response from different areas in WS$_2$/ReSe$_2$ heterostructure.
    (c) Optical image of the WS$_2$/ReSe$_2$ heterostructure. The two monolayers and the bilayer are labeled.
    (d) Illustration of monolayer WS$_2$ that shows the armchair direction (cyan arrow) and the direction of the fundamental laser polarization (black arrow).
    (e) Similar illustration of monolayer ReSe$_2$ that shows the rhenium chain (Re-chain) direction (brown arrow) and the direction of the fundamental laser polarization (black arrow).
    (f) Reflection spectra of the two monolayers (WS$_2$ -- cyan trace, and ReSe$_2$ -- red trace) and the heterobilayer (purple trace).
    (g) SHG spectra measured from different regions shown in (c).
    (h) Log-scale plot of the pump-power dependence of SHG for monolayer WS$_2$ (cyan) and WS$_2$/ReSe$_2$ heterobilayer (purple).
    }
    \label{fig:Fig1}
\end{figure*}

%%%%%%%%%%%%%%%%%%%%%%%%%%%%%%%%%%%%%%%%%%%%%%%%%%%%%%%%%%%%%%%%%%%%%%
\section{Experimental Results}
\subsection{Experimental Setup and Sample Preparation}
A schematic of the SHG beamline is shown in Fig. \ref{fig:Fig1}(a).
Briefly, an output from a near-infrared femtosecond laser (1030 nm, 40 MHz, 200 fs) is focused onto the sample using a near-infrared objective (Plan Apo, 20$\times$, NA 0.40), and the perpendicular component of the reflected signal is detected using a compact spectrometer.
The WS$_2$/ReSe$_2$ sample is placed on a two-axis motorized stage, allowing for spatial mapping of the monolayer and bilayer regions [Fig. \ref{fig:Fig1}(b)].
For the polarization-dependent measurements, we vary the polarization plane of the fundamental laser using a half-wave plate.

The heterobilayer samples are fabricated by isolating and stacking the monolayers of WS$_2$ and ReSe$_2$ using the gold-assisted mechanical exfoliation method. \cite{liu2020disassembling}
A detailed description of the fabrication process is provided in the  \hyperref[Appendix]{\textbf{Appendix A}}.
The samples are annealed at 200$^\circ$C under vacuum for one hour to increase the contact between the two layers, which we find necessary for the enhancement of SHG. 
Figure \ref{fig:Fig1}(c) shows an optical image of a WS$_2$/ReSe$_2$ heterobilayer sample.
We also measure reflection contrast to characterize the monolayer and heterobilayer regions [Fig. \ref{fig:Fig1}(f)], which we will discuss later.

Key properties of WS$_2$ and ReSe$_2$ are summarized below.
The group-VI TMDs, such as MoS$_2$ and WSe$_2$, belong to the $1H$ phase at their equilibrium, and they possess broken-inversion symmetry corresponding to the trigonal prismatic lattice structure.\cite{li2013probing, kumar_second_2013, ribeiro-soares_second_2015}
Their per-layer efficiencies of SHG are known to be higher than those of conventional nonlinear crystals, highlighting their potential for efficient nonlinear optical applications.\cite{li2013probing, wang2019second}
Meanwhile, the group-VII rhenium TMDs, such as ReS$_2$ and ReSe$_2$, belong to the $1T'$ phase at ambient conditions, which is a slightly distorted variant of the $1T$ phase.\cite{jariwala_synthesis_2016}
In the $1T$ phase, each metal atom is surrounded by six chalcogen atoms, forming a symmetric octahedron.
As a result, the $1T'$ phase largely lacks broken-inversion symmetry, and SHG in rhenium TMDs is inherently weak.\cite{akatsuka_180_2024}
We expect that these distinctive crystal symmetries may yield identifiable characteristics in the SHG polarization response, allowing us to distinguish the contributions from individual layers and from hybridization effects.

%%%%%%%%%%%%%%%%%%%%%%%%%%%%%%%%%%%%%%%%%%%%%%%%%%%%%%%%%%%%%%%%%%%%%%
\subsection{SHG Spectra and Spatial Mapping}
\label{Results}
Figure \ref{fig:Fig1}(g) shows SHG spectra obtained from the WS$_2$ monolayer, ReSe$_2$ monolayer, and WS$_2$/ReSe$_2$ heterobilayer.
The heterobilayer has a twist angle of 30$^\circ$ as determined by the polarization dependence in SHG.
The twist angle is defined as the angle between the armchair direction of WS$_2$ and the rhenium chain (Re-chain) direction of ReSe$_2$ [Figs. \ref{fig:Fig1}(d,e)].
For this measurement, the polarization of the incident 1030-nm laser is fixed to maximize the SHG signal from the monolayer regions.
This direction is parallel to the armchair direction of monolayer WS$_2$ and the Re-chain direction of monolayer ReSe$_2$.
The results show that the SHG from the heterobilayer is $49 \pm 1$\% stronger than that from WS$_2$ monolayer, while the SHG from ReSe$_2$ is only 2\% of that of WS$_2$.

\begin{figure}[tb]
    \centering
    \includegraphics[scale=1]{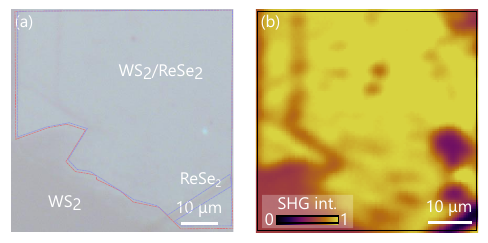}
    \caption{\textbf{SHG mapping of WS$_2$/ReSe$_2$ heterostructure.}
    (a) Optical image of WS$_2$/ReSe$_2$ heterostructure with a twist angle of 30$^\circ$.
    (b) SHG map of the area shown in (a). The results show that the SHG response is flat in the sample area we scanned.
    }
    \label{fig:Fig2}
\end{figure}

Figure \ref{fig:Fig1}(h) shows the laser-power dependence of the SHG signals.
Fittings to the results determine that the power orders are $1.9\pm0.1$ for WS$_2$/ReSe$_2$, and $2.1\pm0.1$ for WS$_2$, indicating that our experiments are performed in the perturbative regime.
Figure \ref{fig:Fig2}(b) shows a mapping result of SHG from the heterobilayer sample.
We observe uniform distributions from the monolayer and heterobilayer regions over the area of tens of microns, ensuring the spatial homogeneity of the enhancement.

These basic SHG measurements [Fig. \ref{fig:Fig1}(d)] already reveal an interesting feature, i.e., the SHG enhancement in the heterobilayer cannot be explained by a simple coherent superposition. \cite{hsu_second_2014, kim_second_2020}
If we assume that the two monolayers are emitting second-harmonic signals independently and the contributions from the two layers are fully in phase, SHG from the heterobilayer would only be $26$\% more than the monolayer WS$_2$.
Therefore, to account for the experimentally observed enhancement, consideration of the interlayer effects is necessary.

%%%%%%%%%%%%%%%%%%%%%%%%%%%%%%%%%%%%%%%%%%%%%%%%%%%%%%%%%%%%%%%%%%%%%%
\subsection{Polarization Dependence}

\begin{figure}[tb]
    \centering
    \includegraphics[scale=1]{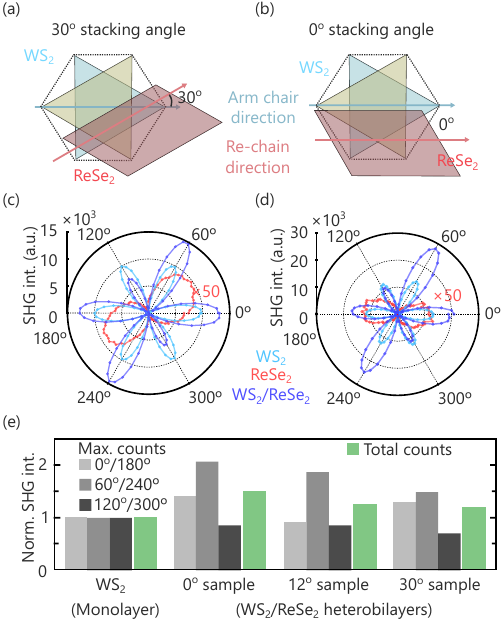}
    \caption{\textbf{Polarization-resolved SHG in WS$_2$/ReSe$_2$ heterostructure and twist-angle-dependence of the SHG enhancement.}
    (a,b) Schematic visualizations of heterostructures with 30$^\circ$ and 0$^\circ$ twist angles.
    (c) Polar plots of the polarization-resolved SHG intensity measured on WS$_2$ (cyan trace) and ReSe$_2$ (red trace) monolayers, and the heterobilayer with a twist angle of 30$^\circ$ (purple trace). The SHG signal from ReSe$_2$ is multiplied by 50 for a better visualization.
    (d) The same plot for a heterobilayer with a twist angle of 0$^\circ$.
    (e) The SHG signals from monolayer WS$_2$ and WS$_2$/ReSe$_2$ heterostructures from individual lobes ($0^\circ/180^\circ$: light gray, $60^\circ/240^\circ$: medium gray, and $120^\circ/300^\circ$: black) are shown in groups. The bars are grouped by sample: monolayer WS$_2$, and WS$_2$/ReSe$_2$ heterobilayers with twist angles of 0$^\circ$, 12$^\circ$, and 30$^\circ$, as indicated categorically on the horizontal axis. The intensity from each lobe is normalized by that of monolayer WS$_2$. Green bars indicate the total SHG counts for WS$_2$ monolayer and each heterostructure, likewise presented as ratios to the monolayer value.
    }
    \label{fig:Fig3}
\end{figure}

To gain more insights into the mechanisms of the SHG enhancement, we fabricate multiple heterobilayer samples and characterize their polarization dependence.
The samples have twist angles at 0$^\circ$, 12$^\circ$, and 30$^\circ$, as illustrated in Figs. \ref{fig:Fig3}(a,b).
The polarization-resolved SHG spectra for the 30$^\circ$ and 0$^\circ$ samples are shown in Figs. \ref{fig:Fig3}(c,d).
For the monolayers, the results show that WS$_2$ exhibits a six-fold symmetric pattern corresponding to their $1H$ phase, whereas ReSe$_2$ exhibits a two-fold symmetric pattern corresponding to their $1T'$ phase.
These results reproduce the known behaviors in the polarization dependence. \cite{janisch_extraordinary_2014, song_extraordinary_2018}

For the 30$^\circ$ heterobilayer sample [Fig. \ref{fig:Fig3}(c)], the four lobes from WS$_2$ that align parallel to the ReSe$_2$ have signals enhanced by $30 \pm 2$\% along 0$^\circ$/180$^\circ$, and by $49 \pm 1$\% along 60$^\circ$/240$^\circ$, whereas the two perpendicular lobes are suppressed by $30\pm2$\% along 120$^\circ$/300$^\circ$, when compared to monolayer WS$_2$ signals.
These results reveal an important fact that SHG is not always enhanced, but it can be suppressed depending on the polarization direction.
This observation further suggests that intensity redistribution within WS$_2$ layer might be taking place on top of the enhancement of the overall SHG intensity, effectively channeling the nonlinear response into selected axes.

For the 0$^\circ$ heterobilayer sample [Fig. \ref{fig:Fig3}(d)], where the armchair direction and the Re-chain direction are parallel, the anisotropic behavior is even more pronounced. 
Note that the $1T'$ phase is two-fold symmetric but lacks mirror symmetry, and thus the heterobilayer as a whole is two-fold symmetric without mirror symmetry.
The SHG signals are enhanced by $101 \pm 1$\% along 60$^\circ$/240$^\circ$, and by $45\pm 1$\% along 0$^\circ$/180$^\circ$, whereas they are suppressed by $12\pm 1$\% along 120$^\circ$/300$^\circ$, relative to monolayer WS$_2$.
Compared to the 30$^\circ$ case, the angular variation in the SHG response is significantly amplified, highlighting the probable influence of relative crystallographic orientation.

Since assessing only the maximum SHG intensities along the lobe directions could obscure possible suppression effects, we examine both the maximum counts and the total SHG intensity to fully characterize SHG modulation in the heterobilayer samples.
Figure \ref{fig:Fig3}(e) summarizes the SHG signals for WS$_2$ and WS$_2$/ReSe$_2$, shown separately for the three lobe-directions and normalized by the response from the monolayer WS$_2$.
Across all heterobilayer samples, the SHG signals from the heterobilayers are enhanced along two of the WS$_2$ armchair directions, and decreased along the third, irrespective of the twist angle between the two monolayers.
Moreover, the total SHG intensity (Fig. \ref{fig:Fig3}(e), green bars) and the maximum SHG counts (Fig. \ref{fig:Fig3}(e), medium gray bars) decrease gradually with increasing twist angle, indicating the influence of interlayer orientation on the observed SHG response.

%%%%%%%%%%%%%%%%%%%%%%%%%%%%%%%%%%%%%%%%%%%%%%%%%%%%%%%%%%%%%%%%%%%%%%
\section{Discussion}

The results so far have shown that van der Waals stacking of monolayer ReSe$_2$, which is SHG inactive on its own, can increase the SHG efficiency of its heterobilayer with WS$_2$, while altering the anisotropic polarization dependence. %; the combined system loses its mirror symmetry.
We can consider three possible scenarios of the underlying mechanisms: (i) originally SHG-inactive ReSe$_2$ becomes SHG-active, or (ii) already SHG-active WS$_2$ becomes more efficient, or (iii) there exist mixed contributions from the two layers.
Given the observation that the polarization dependence of the heterobilayer does not resemble that of the individual monolayers [Figs. \ref{fig:Fig3}(c,d)], it is reasonable to conclude that scenario (iii) is most likely, necessitating us to consider modified electronic structures via interlayer interactions.

We first assess the electronic structure of our samples by measuring their optical reflectivity [Fig. \ref{fig:Fig1}(f)].
The monolayer WS$_2$ shows excitonic features around 2.01 eV, 2.38 eV, and 2.82 eV, corresponding to its A, B, and C excitons, respectively. \cite{Li_measurement_2014}
The monolayer ReSe$_2$, with its bandgap at 1.31 eV, \cite{jariwala_synthesis_2016} remains structureless in the range of our white-light spectrum. %(1.55 - 3.10 eV).
In the heterobilayer, the excitonic features from WS$_2$ exhibit broadening and spectral red shifts, with the shift amounts being $25$ meV for the A exciton, and $30$ meV for the B exciton. 
These results indicate that the electronic structure in the valence and conduction bands is altered by interlayer interactions, with a possible mechanism being charge-density redistribution between the two layers. \cite{ruiz-tijerina_interlayer_2019}

\begin{figure}[tb]
    \centering
    \includegraphics[scale=1]{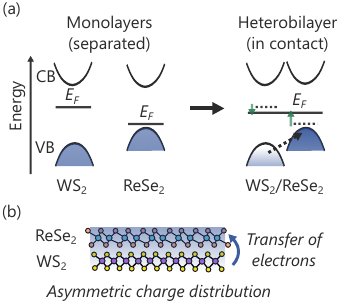}
    \caption{\textbf{Energy band alignments in WS$_2$/ReSe$_2$ heterostructure.}
    (a) Illustration of band alignment and corresponding Fermi energy levels ($E_F$) in two separated monolayers and WS$_2$/ReSe$_2$ heterobilayer. Once the two monolayers are in contact, their Fermi levels shift, as indicated by the green arrows in the right panel, and electrons may flow from the WS$_2$ valence band to the ReSe$_2$ valence band. VB: valence band, and CB: conduction band.
    (b) Schematic of the charge transfer between ReSe$_2$ and WS$_2$, which may induce a permanent shift in the overall polarization of the heterobilayer, as indicated by the color gradient.
    }
    \label{fig:Fig4}
\end{figure}

We then turn to the computational results to verify the band alignment between the two layers, and a schematic illustration is shown in Fig. \ref{fig:Fig4}(a). \cite{han_type-ii_2020, li_type_2018}
According to previous density functional theory calculations, the Fermi levels referenced to the vacuum level are $-5.03$ eV and $-4.55$ eV for ReSe$_2$ and WS$_2$ monolayers, respectively. \cite{hu_atomic_2024}
Therefore, when the two layers are in contact, electrons transfer from WS$_2$ to ReSe$_2$ for their Fermi levels to shift to the same level, forming an out-of-plane dipole moment [Fig. \ref{fig:Fig4}(b)].
This transfer of electrons modifies the band structure in the vicinity of the upper valence and the lower conduction bands of the heterostructure.
The observed red shifts in the WS$_2$ excitons agree with this picture, reflecting the occurrence of effective hole doping in WS$_2$ [Fig. \ref{fig:Fig1}(f)]. \cite{wang_interlayer_2016, qin_growth_2019, guan_atomic_2024}

%%%%%%%%%%%%%%%%%%%%%%%%%%%%%%%%%%%%%%%%%%%%%%%%%%%%%%%%%%%%%%%%%%%%%%
\section{Conclusions}
\label{Conclusions}
The following conclusions may be drawn from the results obtained in this work.
First, an interlayer charge redistribution is likely occurring in our samples, and the induced out-of-plane dipole moment can be an important factor for the observed SHG enhancement [Fig. \ref{fig:Fig4}(b)].
Moreover, the charge redistribution is likely to be entangled with the renormalization of the valence and conduction bands induced by the modified carrier distribution.
These mechanisms are suggested by the observed peak shifts in the reflectivity measurements [Fig. \ref{fig:Fig1}(g)].
Second, the observed anisotropy or the twist-angle dependence (Fig. \ref{fig:Fig3}) cannot be explained by the band alignment alone, as such models do not account for detailed crystal structures.
Instead, it is necessary to consider the formation of interlayer hybrid bands, moving beyond the simple charge-transfer picture.
The sensitivity of anisotropic SHG response to interlayer effects observed in this study is distinctive compared to those typically observed in group-IV TMD heterobilayers (e.g., WS$_2$/MoSe$_2$ \cite{kim_exciton-sensitized_2023}).
We attribute this to the different crystal phases adopted by WS$_2$ and ReSe$_2$, whose contributions can be discerned in polarization-resolved SHG measurements.
Third, although our analysis has focused on SHG enhancement, the effect is highly anisotropic and can even lead to suppression at certain crystal angles [Fig. \ref{fig:Fig3}(e)].
This points to a possible role of intensity-borrowing mechanisms, \cite{Orlandi73} which may be induced within the SHG-active WS$_2$ through interaction with ReSe$_2$.

Looking forward, this study highlights the promising potential of van der Waals nonlinear optics for realizing tunable platforms that can be controlled not only in terms of intensity, but also polarization dependence. 
Furthermore, our work suggests that the sensitivity of SHG to interlayer interactions can be significantly strengthened when different crystal phases are adopted in the composite layers.

%%%%%%%%%%%%%%%%%%%%%%%%%%%%%%%%%%%%%%%%%%%%%%%%%%%%%%%%%%%%%%%%%%%%%%%%%%%%%%%%%
\begin{acknowledgments}
K.S., T.Y., Z.Z., and Y.K. acknowledge the startup funds from the University of Michigan.
Q.L. and H.D. acknowledge the support by the Army Research Office (W911NF-25-1-0055) and the Betty and Gordon Moore Foundation (GBMF10694).  
A.C.J. acknowledges support from the Stanford Tomkat fellowship.
The preparation of the samples is supported by the Defense Advanced Research Projects Agency (DARPA) under Agreement No. HR00112390108.
The preparation of the gold thin films was performed at the University of Michigan Lurie Nanofabrication Facility.
\end{acknowledgments}

%%%%%%%%%%%%%%%%%%%%%%%%%%%%%%%%%%%%%%%%%%%%%%%%%%%%%%%%%%%%%%%%%%%%%%%%%%%%%%%%%
\section*{Author Declarations}
\subsection*{Conflict of Interest}
The authors declare no conflicts of interest.

\subsection*{Author Contributions}
K.S., T.Y., and Z.Z. performed the experiments and analyzed the results.
K.S., T.Y., Z.Z., Q.L., and A.C.J. fabricated the samples.
H.D., F.L., and Y.K. supervised the project.
All authors contributed to the preparation of the manuscript.

%%%%%%%%%%%%%%%%%%%%%%%%%%%%%%%%%%%%%%%%%%%%%%%%%%%%%%%%%%%%%%%%%%%%%%%%%%%%%%%%%
\section*{Data availability}
The experimental data presented in this study are available from the corresponding author upon request. 

%%%%%%%%%%%%%%%%%%%%%%%%%%%%%%%%%%%%%%%%%%%%%%%%%%%%%%%%%%%%%%%%%%%%%%%%%%%%%%%%%
\section*{Appendix}
\label{Appendix}
\subsection{Gold-tape Exfoliation Method to Make Monolayers and Heterobilayers}
Bulk crystals of ReSe$_2$ and WS$_2$ are purchased from HQ Graphene.
We use the gold-tape exfoliation method to prepare the samples on SiO$_2$ substrates (University Wafer, Inc.). \cite{liu2020disassembling}
A 150-nm-thick gold layer is deposited on a silicon wafer (University Wafer, Inc.) using an electron beam evaporator.
A 10\% wt poly-vinylpyrrolidone (PVP) solution (Thermo Scientific, M.W. 40,000) in a 1:1 mixture of ethanol and acetonitrile is spin-coated on the gold film as a protecting layer.
To prepare the gold tape, we apply a thermal-release tape (Semiconductor Corp., $90^\circ$C release temperature) on the PVP-coated gold.
The gold tape is then gently pressed onto a freshly cleaved WS$_2$ crystal.
When the tape is lifted off, it takes a mm-scale monolayer attached to the gold surface, which is then transferred onto the SiO$_2$ substrate.
The thermal-release tape is removed by heating the assembly at $135^\circ$C, while the PVP is removed by dissolving in water for 4 hours and in acetone for 30 minutes.
The gold layer is dissolved in an aqueous solution of potassium iodide and iodine (2.5 g iodine, 10 g KI in 100 mL deionized water; iodine, crystalline, Thermo Scientific, 99.5\%; potassium iodide, Thermo Scientific, 99\%).
The monolayer on the substrate is then rinsed with deionized water and isopropanol and dried with nitrogen.

To prepare the heterobilayers, the substrate is placed on a transfer stage, and a monolayer of ReSe$_2$ is exfoliated using gold tape. 
The edges of the pre-prepared monolayer on the substrate and the freshly peeled monolayer on the gold tape are matched to have the desired stacking angle between the monolayers.
Once the target angle is reached, they are brought closer to make van der Waals heterostructures. 
The thermal release tape, PVP, and gold are removed following the processes as described before. \cite{johnson_hidden_2024}
The actual stacking angles of the bilayers are measured by performing polarization-resolved SHG measurements.
We anneal the exfoliated samples at 200$^\circ$C under vacuum in a tube furnace to increase the coupling strengths between the layers.

\subsection{SHG Polarization Measurements}
The crystal orientations of WS$_2$ and ReSe$_2$ monolayers are determined by SHG polarization measurements.
A linearly polarized femtosecond laser light (1030 nm, 40 MHz, 200 fs) is focused onto the monolayer sample with a microscope objective (Plan Apo NIR, 20$\times$, NA 0.40). 
The reflected SHG signal at 515 nm is passed through a short-pass filter and a linear polarizer and detected using a compact spectrometer (USB 2000+).
To obtain the polarization-dependent SHG signal, we rotate the plane of the fundamental laser polarization using a half-wave plate on a motorized rotational mount.
During the experiments, we select the perpendicular component of the generated second harmonic by placing a polarizer on another motorized rotational stage.
We combine an optical microscope with the SHG setup to selectively pump the area of interest.
For the monolayers, when the laser polarization is parallel to the crystal axis (i.e., the armchair direction for WS$_2$ and the Re-chain direction for ReSe$_2$), the maximum SHG is obtained, and the intensity changes sinusoidally with the angle between the laser polarization and the crystal axis.
We define the twist angle as the angle between the two crystal axes of the two monolayers, with an uncertainty of $\pm1^\circ$.

\subsection{Reflection Contrast Measurements}
The room-temperature steady-state reflection contrast measurements are carried out using a home-built reflectometer setup equipped with a white light source.
The setup is also combined with an optical microscope.
We focus the white light onto the desired area on the sample using an objective lens (L Plan, 50$\times$, NA 0.55). 
The reflected signal from the sample is collected through the objective and refocused with an achromatic lens to a compact spectrometer.
During the measurements, the spectrometer has been set to acquire spectra for a total period of 5 seconds.
We plot $\text{log} \left(R/R_0 \right)$ vs photon energy (eV) to get the static reflection contrast spectra of the monolayers and heterobilayer, where $R$ and $R_0$ are the reflectances from the sample on the substrate and the bare substrate.

%%%%%%%%%%%%%%%%%%%%%%%%%%%%%%%%%%%%%%%%%%%%%%%%%%%%%%%%%%%%%%%%%%%%%%%%%%%
%References
\section*{References}
\bibliographystyle{apsrev4-2.bst} 
\bibliography{Bibliography}

%%%%%%%%%%%%%%%%%%%%%%%%%%%%%%%%%%%%%%%%%%%%%%%%%%%%%%%%%%%%%%%%%%%%%%%%%%%
\end{document}